# Enhanced stochasticity in irradiated vanadium oxide oscillators

Nareg Ghazikhanian*[1,2], David J. Alspaugh[1], Pavel Salev[3], Lorenzo Fratino[4], Marcelo Rozenberg[5], and Ivan K. Schuller[1,2]

1. Department of Physics, University of California, San Diego

2. Materials Science and Engineering Program, University of California, San Diego

3. Department Physics and Astronomy, University of Denver

4. CY Cergy Paris Université, CNRS, Laboratoire de Physique Théorique et Modélisation

5. Université Paris-Saclay, CNRS, Laboratoire de Physique des Solides

*Corresponding author : nghazikh@ucsd.edu

**Abstract**

Insulator-to-metal transition materials are highly sensitive to even minute deviations of stoichiometry, lattice defects, and disorder, which provides opportunities to engineer their electrical switching characteristics. Using $V_2O_3$ as a prototypical metal-insulator transition resistive switching material, we demonstrate that localized focused ion beam irradiation can induce stochastic oscillatory dynamics in simple two-terminal switching devices. After irradiating the material, we observed an unusual dynamic regime where the voltage induced metallic state momentarily collapses into an insulating state, which results in a rapid current flickering that is qualitatively different from the conventional current spiking in a Pearson-Anson type oscillatory circuit implemented using the pristine material. Furthermore, the current flickering timing in the irradiated devices becomes progressively more random and more sparse with increasing input voltage, resulting in nonlinear and nondeterministic oscillatory behavior. The irradiation also leads to a dramatic reduction in switching power required to induce the current oscillations. These results are elucidated through random resistor network simulations which indicate that a small number of local sites can control the electrical metal-insulator transition switching properties in large devices with high defect concentration. Our results show that selective focused ion beam irradiation provides exciting prospects for engineering and tuning novel stochastic behaviors in emergent technologies that rely on the intrinsic randomness of physical processes.

## Introduction

Intrinsic dynamic stochasticity of material properties can be leveraged for a wide range of electronics applications including probabilistic p-bits[1-5], random number generators,[4, 6-9] and beyond-CMOS elements in neuromorphic computers[10-15]. Metal-insulator transition (MIT) resistive switching (RS) materials often show intrinsic switching stochasticity and can offer multiple practical advantages including the ability to manipulate their resistivities quickly, repeatedly, and efficiently via multiple distinct external stimuli.[16, 17] Under the application of electrical biasing, these materials undergo an electronic phase transition which results in an orders-of-magnitude change of their resistivity. This electrical switching typically occurs by the formation of a metallic phase filament percolating though the insulating phase matrix.[18-24] Local morphology and stoichiometry often dictate the phase transition characteristics, including the electronic phase transition temperature ($T_c$), switching power, optical characteristics, strain distribution, and filament geometry.[20, 21, 25] Focused ion beam irradiation has been demonstrated to be an efficient method of locally modifying defect concentrations, which can result in significant changes in both equilibrium and non-equilibrium transport properties[26-28]. Here, we employ focused ion beam irradiation to tune the RS stochasticity in MIT materials.

As local defect distribution can induce significant changes in the local electronic landscape, it is feasible that these changes can result in qualitatively different resistive switching dynamics. Defects are expected to create local variations of $T_c$ and carrier concentration, which impact the resistivities of insulating and metallic phases.[29, 30] It has been shown that the resistivity ratio between the insulating and metallic states, $\rho_{insulator}/\rho_{metal}$, determines the rate and width of metallic filament formation during the electrical switching, where higher ratios lead to faster incubation times and narrower filaments.[22, 31, 32] Defects can also produce randomly distributed filament nucleation sites leading to highly heterogenous current flow. Modifying the local defect concentration and its spatial distribution, thus, can have a significant impact on the filament formation and relaxation timescales and on the filament morphology, leading to qualitatively different RS dynamics.

In this work, we demonstrate that focused ion beam (FIB) irradiation induces novel stochastic behaviors in Pearson-Anson type oscillators based on MIT RS materials. We used $V_2O_3$, a prototypical Mott insulator, because it has been demonstrated that electronic properties of $V_2O_3$ are extremely susceptible to irradiation effects, making it an ideal material for this experiment.[27, 28] The irradiated $V_2O_3$ devices develop an unusual stochastic regime in their oscillatory behavior under the application of a dc voltage. In this stochastic oscillation regime, the metallic state persists for significantly longer duration than the spiking oscillations in pristine material, resulting in an anti-spike-like current flickering with progressively sparser and more random inter-spike periodicity as the driving dc voltage increases. After ion irradiation, the devices require significantly less power and smaller parallel capacitance to sustain oscillations. This demonstrates that selective ion irradiation is an efficient tool for inducing stochasticity in RS materials, an exciting prospect for engineering novel, scalable, and energy-efficient electronics that leverage stochasticity of physical processes.

## Results

We first explored the equilibrium transport properties of two-terminal $V_2O_3$ devices before and after selective $Ga^+$ ion beam irradiation. We used a 30 keV focused $Ga^+$ ion beam with the dose of $6.2 \cdot 10^{14}$ Ga ions/cm$^2$ to create a 1-µm-wide irradiated region spanning along the 10-µm-wide device defined on a 10-nm-thick $V_2O_3$ thin film (Fig. 1A). Transport of Ions in Matter[33] (TRIM) simulations (Fig. 1B) indicated that a single Ga ion induces ~200 atomic vacancies in the $V_2O_3$ layer, leaving a path of atomic displacements throughout the entire film thickness. Because the simulations showed that the majority of the ions pass through the film and implant into the substrate and there are orders of magnitude more vacancies than incident ions, atomic vacancies are likely the dominant defects at play in our samples. Figure 1C compares the resistance vs. temperature of a device before (blue line) and after (red line) local $Ga^+$ irradiation. Several qualitatively new features emerge after the $Ga^+$ bombardment. First, the resistance-temperature curve appears to be split into two parts (blue and pink in Fig 1C) because of the simultaneous presence of pristine and irradiated regions inside the device. The shape of the resistance-temperature curve can be understood by considering that the $Ga^+$ irradiation locally lowers $T_c$ of $V_2O_3$, and the pristine and irradiated regions act as two resistors in parallel. Second, the resistance of the high-temperature metallic state increases while the resistance of low-temperature insulating state decreases after the irradiation. Third, the irradiated device shows sudden resistance jumps of up to 2 orders of magnitude (similar to confined $V_2O_3$ nanowires.[34]) above a 5 K temperature variance upon repeated thermal cycling (labeled "Stochastic hysteretic regime" in pink Fig. 1C) In contrast, the pristine device shows little to no variability during consecutive thermal cycles (labeled "Pristine hysteretic regime" in Fig. 1C). The Ga+ irradiation, therefore, changes local $T_c$, $\rho_{\text{insulator}}/\rho_{\text{metal}}$ ratio, and introduces a stochastic behavior evident even in the equilibrium properties.

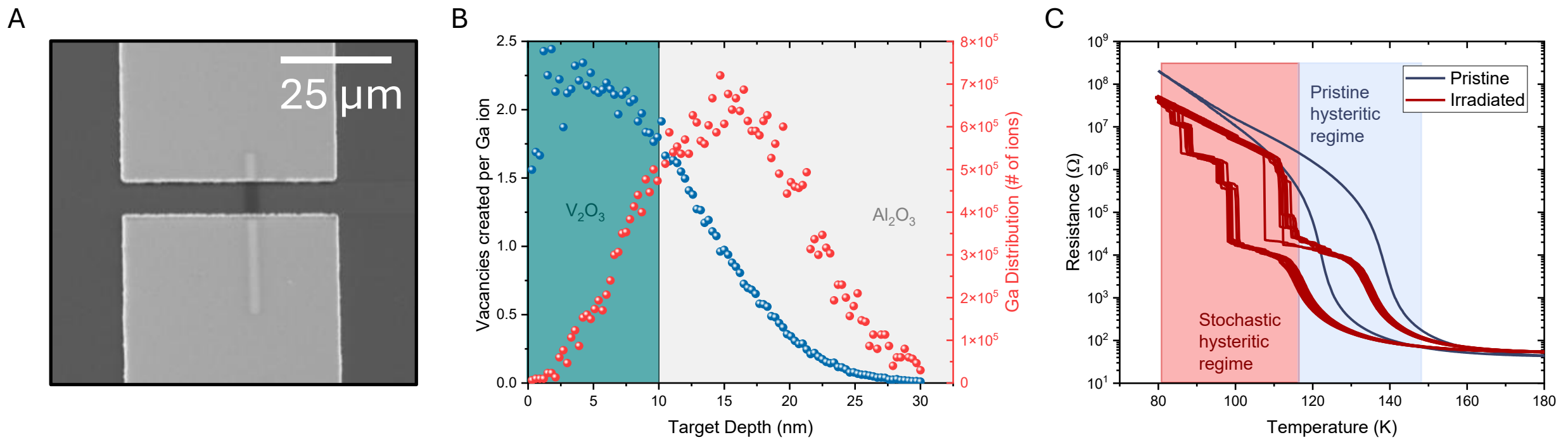

**Figure 1. A)** Scanning electron micrograph of a $V_2O_3$ device irradiated with a 1-µm-wide strip of $Ga^+$ ions. **B)** Tracking Range of Ions in Matter (TRIM) simulation data for 30 keV $Ga^+$ ions bombarding a 10 nm $V_2O_3$ film on an $Al_2O_3$ substrate. While the vacancy distribution (blue curve) created per incident ion is relatively consistent throughout the entirety of the film depth, the Ga ion distribution (red curve) indicates that most incident ions are implanted into the substrate. Integration of the blue curve indicates that ~200 vacancies are created in $V_2O_3$ per incident ion, suggesting that vacancies are the dominant type of defect. **C)** Multiple cycles of resistance vs. temperature measurements for a $V_2O_3$ device before (blue line) and after (red line) irradiation. The irradiated device displays significant cycle-to-cycle variability and large resistance jumps. The contribution of the irradiated channel leads to a splitting of resistance-temperature curve into two branches due to a depressed $T_c$ in the locally irradiated regions.

The ion beam irradiation of $V_2O_3$ develops novel electrical spiking behavior when incorporated into Pearson-Anson type oscillator circuits (Fig. 2). As it is common among many MIT materials, when connecting a two-terminal device to a resistor in series and to a capacitor in parallel (Fig. 2F), the device develops self-oscillations upon the application of an above-threshold dc voltage.[32, 35] This "spiking" is attributed the periodic formation and disruption of a metallic phase filament governed by the electrical dynamics in the RC circuit and thermal interactions between the MIT film and underlying substrate.[22, 35] Such oscillations develop in our pristine $V_2O_3$ devices at 95 K using a 100 kΩ series resistor and 1 nF parallel capacitor (Fig. 2A). Inducing the spiking oscillations in the pristine devices required the application of a relatively large dc voltage (>30 V) because of the large device dimensions ($10\times40$ $\mu m^2$). As applied voltage increases, we observed a nearly linear spiking frequency increase (Fig. 2C, black line). This linear frequency increase trend continues until the upper threshold voltage of ~200 V above which the device remains in the persistent switched state. Overall, the pristine $V_2O_3$ exhibits typical spiking oscillations expected of an MIT RS device.

A qualitatively different oscillatory behavior is found in the irradiated device (Fig. 2B). In these measurements, we used the same series resistor (100 kΩ) but without the need for a parallel capacitor. Possibly, the parasitic capacitance is sufficient to sustain oscillations. While the pristine device required a minimum of 30 V to induce the current spikes of ~24 mA amplitude, the same device after irradiation had the oscillation threshold of only 3.47 V and the current spike amplitude of ~8 μA, corresponding to a factor of ~25,000 reduction of power per oscillation. Importantly, the shape of the current oscillations in the irradiated device is different compared to the pristine device. In the pristine device (Fig. 2A), the current spikes are extremely sharp, resulting in a small oscillation duty cycle (e.g. Fig. 2E) that increases very minimally (not exceeding 10%) with increasing dc driving voltage throughout the entire oscillatory range (Fig. 2C, blue line). In the irradiated device (Fig. 2B), the current oscillations resemble the oscillations in the pristine device only at low dc driving voltages (<4.5 V). When the driving voltage increases, the current spikes become broader. At high driving voltages, the current remains high for almost the entire oscillation period and only briefly flickers off producing inverted spikes (for example, the current trace at 7.82 V in Fig. 2B). This broadening of the shape of the oscillation features is further emphasized by the steadily increasing oscillation duty cycle with increasing driving voltage, from ~10% at 3.47 V to ~90% at 8.57 V (Fig. 2D, blue line). The observed spike broadening indicates that the $Ga^+$ irradiation changed the MIT switching dynamics in $V_2O_3$ samples.

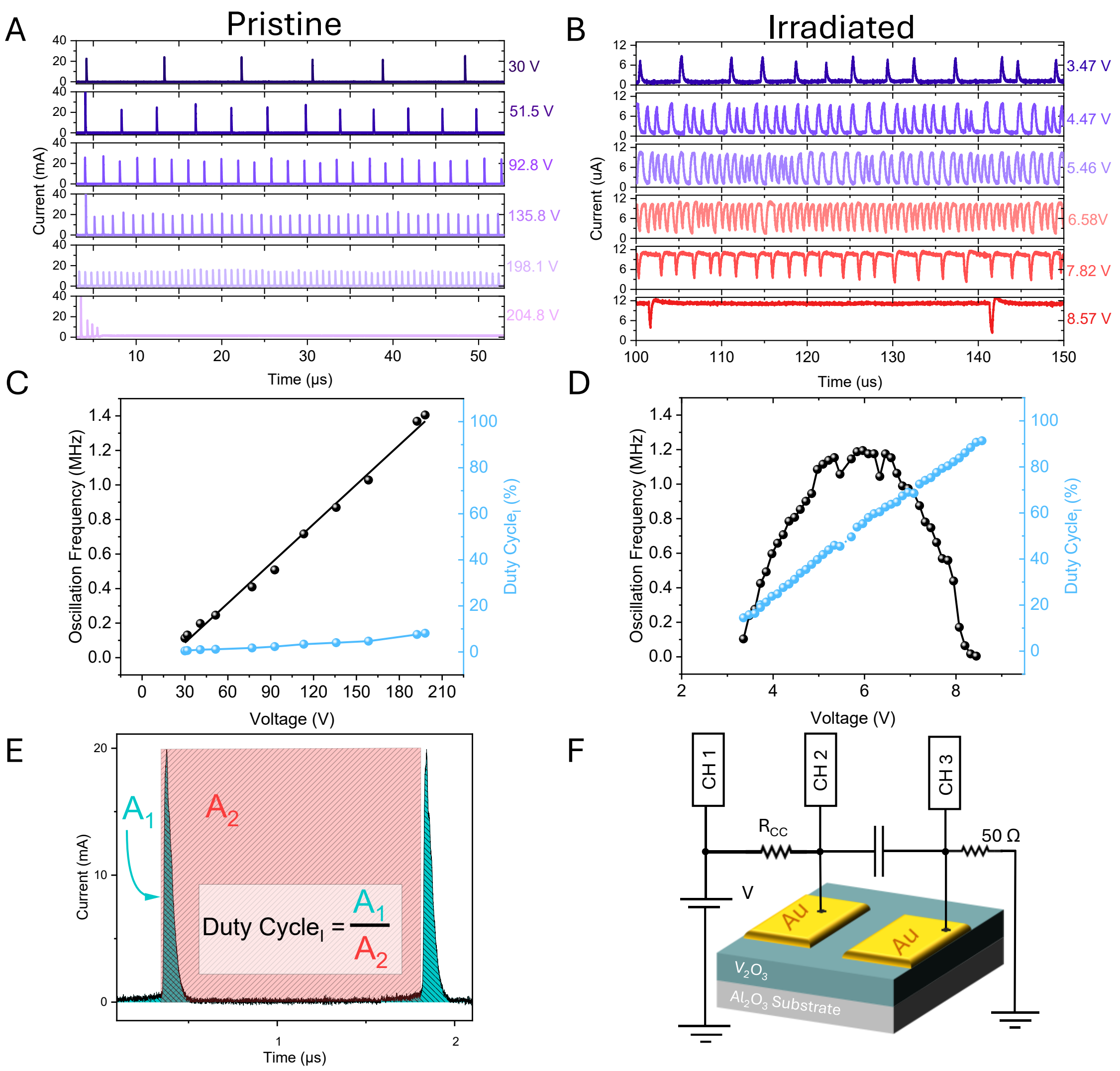

**Figure 2. A)** Current time traces at several dc driving voltages recorded in the pristine $V_2O_3$ device showing highly periodic and sharp spiking. **B)** Current time traces at several dc driving voltages recorded in the irradiated device. The oscillation behavior becomes noticeably more stochastic compared to the pristine device. Current spike broadening can be also observed. **C, D)** Oscillation frequency (black line) and duty cycle (blue line) vs. driving dc voltage for the pristine (C) and irradiated device (D). Irradiated device exhibits a nonmonotonic frequency-voltage dependence and a large increase of the duty cycle with increasing the dc driving voltage. **E)** Definition of the duty cycle used to obtain data shown in panels C) and D). **F)** Schematic of the measurement circuit to probe the oscillatory dynamics. We note that a 1 nF external capacitor was used in the pristine device measurements, while the irradiated device measurements relied only on the parasitic capacitance without connecting an external parallel capacitor.

The irradiated devices also exhibited a qualitatively different oscillation frequency-dependence on voltage as compared to the pristine devices. As discussed above, the spiking frequency steadily increases with increasing driving dc voltage in the pristine device (Fig. 2C, black line). After $Ga^+$ irradiation, the frequency-voltage dependence becomes nonmonotonic (Fig. 2D, black line): the frequency first increases from ~0.1 MHz to ~1.2 MHz in ~3.5 - 5.5 V range, but then decreases to 4.1 kHz as the dc driving voltage is increased further to ~8.5 V. We note that although the oscillation measurements in the irradiated devices were performed without the use of a parallel capacitor, the measured frequency range in the irradiated and pristine devices were comparable, even though the RC constant of the irradiated device circuit must be at least several orders of magnitude smaller compared to the pristine device circuit. The fact that the RC constant appears to not be a dominant factor determining the oscillation frequency and the observed nonmonotonic frequency-voltage dependence provide further evidence that the $Ga^+$ irradiation changed the MIT switching dynamics of $V_2O_3$.

Interestingly, the oscillations in the irradiated devices exhibited very high stochasticity, i.e., a high degree of randomness of the inter-spike intervals. This increased stochasticity can be directly seen by comparing the current traces in Figure 2. While the current spikes in pristine device appear to be periodic (Fig. 2A), the oscillation period in the irradiated devices exhibit large cycle-to-cycle variations. For quantitative comparison of stochastic properties, we measured the self-oscillations' dynamics in pristine and irradiated devices under the same conditions, i.e., using the same series resistor of 100 kΩ and same parallel capacitor of 1 nF (Fig. 3), in contrast to the results in Figure 2 where different series capacitances were used. We also show in Figure 3 the time traces of voltage across the switching device to provide an alternative perspective on switching dynamics. Similar to the results in Figure 2, we observed a nonmonotonic frequency-voltage dependence in the irradiated device (Fig. 3D, black line). The maximum frequency, however, was reduced by a factor of ~20 due to the presence of the 1 nF parallel capacitance, indicating that the dynamics in these experiments is governed by the RC constant of the circuit. To quantify stochasticity, we analyze the statistics of the inter-spike periods. In the pristine device (Fig. 3C, orange line), the standard deviation of the inter-spike period distribution remains low (below 10%) and rapidly decreases with increasing the dc driving voltage. In contrast, the inter-spike period standard deviation in the irradiated device is noticeably larger and it steadily increases with increasing dc driving voltage, even exceeding 100% at 2.3 V (Fig. 3D, orange line). Therefore, we conclude again that the irradiated devices exhibit qualitatively different electrical MIT switching dynamics characterized by stochastic alterations between the insulating and metallic phases as compared to the nearly deterministic oscillations in pristine $V_2O_3$.

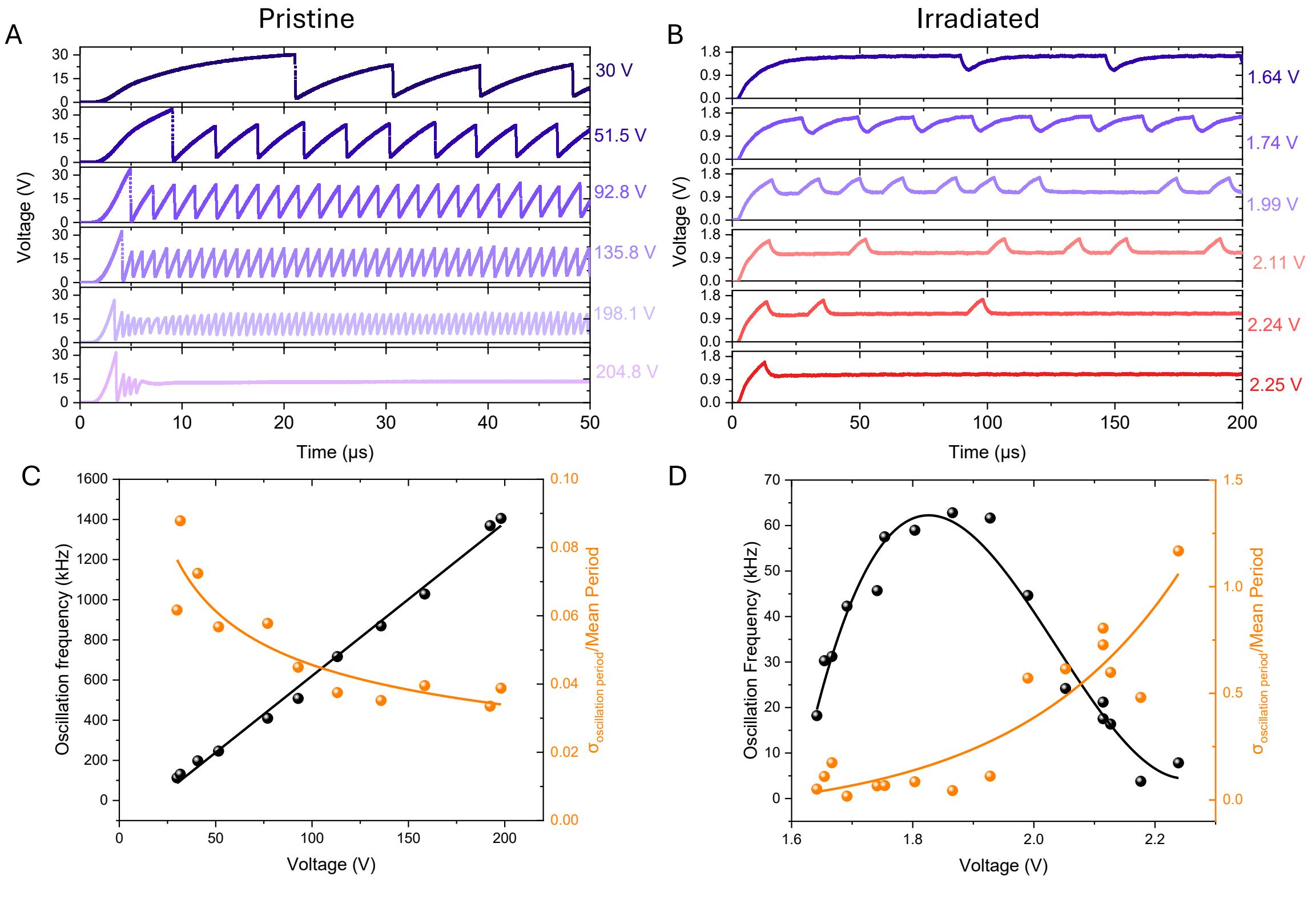

**Figure 3. A, B)** Voltage time traces showing the oscillation dynamics in pristine (A) and irradiated (B) devices. Both sets of measurements were made using the same 100 kΩ series resistor and 1 nF parallel capacitor, which results in much slower oscillations in the irradiated device compared to data shown in Fig. 2. The voltage oscillations in the irradiated device display noticeable stochasticity and much smaller amplitude, suggesting the metal-insulator phase oscillations are confined to the irradiated region. **C, D)** Oscillation frequency (black line) and frequency standard deviation (orange line) vs. input voltage for the pristine (C) and irradiated (D) devices. Even though the 1 nF series capacitor drastically slows down the oscillations in the irradiated device, a much larger frequency standard deviation (i.e., much higher oscillation stochasticity) can be observed in the irradiated device compared to the pristine device.

The voltage traces shown in Figure 3 also provide an additional insight into the nature of filamentary confinement. The pristine device shows a large voltage oscillation amplitude, and after each spiking event, the voltage drops almost to zero, indicating a wide switched region (i.e. small resistance during the spike). The irradiated sample has a much smaller amplitude of voltage oscillation. Furthermore, the voltage across the irradiated device never drops near zero, indicating that the device remains in a comparatively high resistance state throughout the oscillation cycle. Considering that the irradiated region was only 1 μm wide compared to the 40 μm width of the device, this voltage amplitude behavior suggests that the phase oscillation area remains geometrically confined to the irradiated region, while the pristine $V_2O_3$ regions surrounding the irradiated channel likely do not contribute to the oscillatory behavior.

As shown here (Fig. 1) and reported previously[26, 36], ion beam irradiation affects both the resistivity ratio $\rho_{\text{insulator}}/\rho_{\text{metal}}$ and the phase transition temperature $T_c$. In order to understand how changing $\rho_{\text{insulator}}/\rho_{\text{metal}}$ and $T_c$ affects the oscillatory behavior of RS devices we simulated the electrical dynamics of the Pearson-Anson circuit using a resistor network with mesoscopic-scale disorder that replicates selective ion beam irradiation.[22, 37] As has been shown in previous works, resistor network models are capable of reproducing sustained voltage oscillations under a dc driving voltage[35]. We used a network of 100×106 resistors with each resistor having a local temperature dependent transition probability between $\rho_{\text{insulator}}$ to $\rho_{\text{metal}}$ states to emulate device spicing (Fig 4). The $\rho_{\text{insulator}}$, $\rho_{\text{metal}}$, and $T_c$ of each resistor were chosen to reproduce the pristine oscillation behavior shown in Figure 1C. The dynamics of the model is determined via coupled time-dependent electrical and thermal equations that have been described elsewhere[22]. To model the localized disorder due to the $Ga^+$ ion beam irradiation, we modified the resistor network such that a random pattern of the resistors within a 10-cell wide column connecting the two electrodes have $\rho'_{\text{insulator}} < \rho_{\text{insulator}}$, $\rho'_{\text{metal}} > \rho_{\text{metal}}$, and $T'_c < T_c$. At the center of the column, we modified 70% of the resistors, while we changed only 35% of resistors towards the edges of the column to mimic the expected spatial defect distribution in experiments. We stress that the modified resistors representing the local defect sites had a similar metal-insulator transition probability as the resistors representing the pristine material, i.e., the results described below do not stem from artificially introducing an enhanced stochasticity into the model.

Figure 4A and 4C show the simulated current time traces, oscillation frequency, and the frequency's standard deviation of the pristine device. Similar to the experimental results, the simulations (Fig. 4C) show a monotonic increase of the oscillations' frequency (black dots) and a monotonic decrease in the frequency's standard deviation (orange dots) with increasing dc driving voltage. The presence of small stochasticity, i.e., non-zero standard deviation, is due to the fact that each resistor in the network transitions between insulating and metallic states with a given probability. Because all resistors are identical in the pristine device, the stochastic transitioning character associated with individual resistors is averaged out when a large number of resistors undergo the transition during each oscillation cycle.

Figure 4B and 4D show the current traces, oscillation frequency, and the frequency's standard deviation obtained from the simulations featuring the irradiation-induced defects. Here, we found a markedly different behavior compared to the simulations of the pristine material. First, the oscillatory behavior emerges at significantly lower input voltages, about an order of magnitude less than those of the pristine case, and the oscillations become extremely stochastic, as can be seen both the in the oscillation shape and period. Like the experimental results, the simulations showed a sharp increase of the frequency's standard deviation as the driving dc voltage increases (orange dots). The simulations also reproduced the oscillation frequency decrease as the oscillations become more stochastic with increasing input voltage, resulting in a nonmonotonic frequency-voltage dependence (black dots).

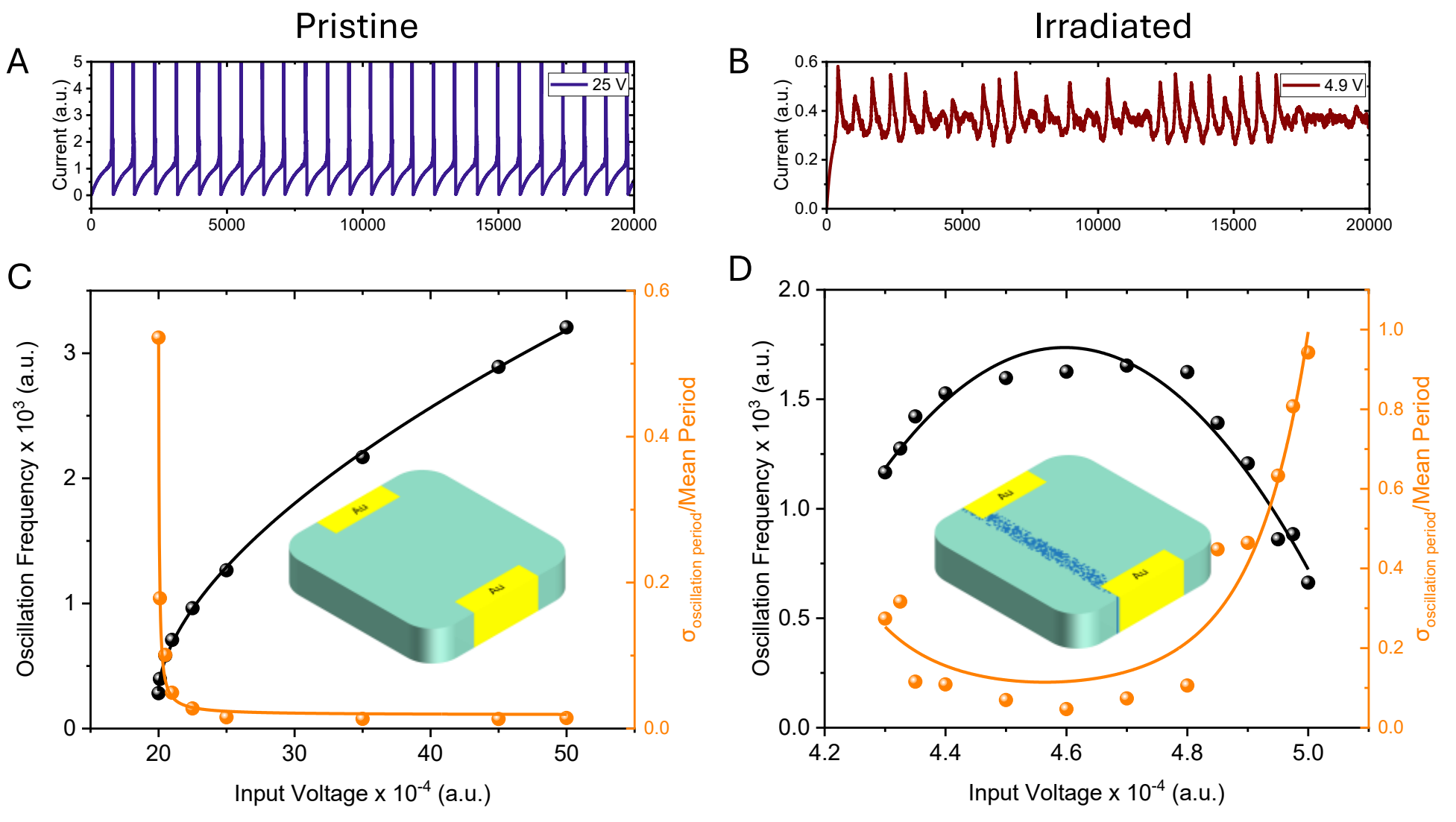

**Figure 4. A, B)** Simulated current spike traces in pristine (A) and irradiated (B) devices. A stark contrast between the highly periodic and sharp spikes in the pristine device vs. stochastic broad spikes in the irradiated device can be observed. **C, D)** Simulated frequency (black line) and frequency standard deviation (orange line) vs. input voltage in pristine (C) and irradiated (D) devices. Simulation results are in a good qualitative agreement with the experimental observation in Fig. 3. The inset schematics illustrate the distribution of defects (blue spots) in the resistor network model.

The oscillation dynamics observed in simulations of the irradiated device, including the nonmonotonic frequency-voltage dependence and enhanced stochasticity can be explained via the formation and dissolution of metallic filaments through the region containing both pristine and defect sites. Defect sites have lower $T_c$ and lower insulating state resistivity. As a result, defects have lower electrical switching threshold, and the metallic filament initially forms along the randomly placed defect resistors. At input voltages above ~$4.8\times10^4$ a. u., these defect-formed filaments start to exhibit significantly longer lifetimes, which explains the oscillation frequency decrease. The defect resistor density, however, is insufficient to form a continuous filament through the defects only, thus the filament also passes through several isolated pristine resistors. Because pristine resistors have higher $T_c$, the isolated pristine resistors inside the defect resistor filament act as weak links. The ON-current associated with the percolating filament can be interrupted even if one of these weak-link pristine resistors briefly flickers off, which is a stochastic process in our model. While in the pristine device simulations, the small stochasticity due to the transition probability of individual resistors averages out because many identical resistors undergo the transition, in the irradiated device simulations, the switching is controlled by a small number of weak links. The stochasticity of a few individual weak-link sites, therefore, controls the behavior of the entire device, resulting in a strongly increased oscillation randomness, like the experimental observations.

## Discussion

The resistor network simulations are in qualitative agreement with the experimental observations: after the irradiation, the devices require smaller voltage to induce oscillations, the frequency-voltage dependence is nonmonotonic, and the oscillations exhibit enhanced stochasticity. The concept of the weak links, which is central to explain the emergence of the enhanced stochasticity, however, remains a plausible hypothesis. Testing this hypothesis requires performing advanced spatiotemporal imaging of the oscillations in pristine and irradiated $V_2O_3$ devices. Such imaging will require nanoscale spatial resolution to track the phase flipping of individual domains and a high resolution in the time domain to capture the oscillation dynamics. Utilizing higher capacitance in the Pearson-Anson circuit can lower the oscillations frequency substantially (as we show in Fig. 2 and 3), however, the nanoscale phase imaging even at kHz frequencies presents a substantial experimental challenge.

The resistor network model takes into account only the thermal effects (Joule heating), which is a natural assumption considering that $V_2O_3$ before and after irradiation exhibit a thermally driven MIT (Fig. 1). Previous works, however, provided evidence that focused ion beam irradiation can induce a change in the resistive switching mechanism in vanadium oxides from one driven by Joule heating to one dominated by electric-field carrier generation[37, 38]. However, it remains unclear the extent that the heating vs. electric field effects play a role in the resistive switching process for the irradiated devices. In our experiments, we observed a factor of ~25,000 reduction of the power per spike, which is not quantitatively reproduced in the electrothermal simulations. This suggests that the role of defects might extend beyond simple current focusing and localizing Joule heating, which promotes the filament formation in the region of high defect concentration. Extending the resistor network model to account for both thermal and non-thermal effects (e.g., electrostatic carrier generation assisting the collapse of the insulting state) may provide further insights into the changes of oscillation dynamics induced by the local ion irradiation in MIT switching materials.

## Conclusion

We showed that a small dose of focused $Ga^+$ ion irradiation in $V_2O_3$ resistive switching devices introduces stochastic oscillatory dynamics when implemented in Pearson-Anson type circuits. In the irradiated devices, we observed significant cycle-to-cycle deviations of the electrical spiking period, which was also accompanied by non-monotonic frequency-voltage dependence and broadening of individual current oscillations ultimately resulting in the appearance of an unusual, inverted spike shape when the driving dc voltage increases. Our resistor network simulations imply that the enhanced stochasticity arises from a small number of isolated pristine sites acting as weak links that can break the metallic filament percolating thorough the highly defected areas. As changes of equilibrium transport and electrical switching properties have been reported in multiple ion-irradiated RS systems[39-41], it is feasible that the methodology to engineer stochastic properties described in this work can also extend to other materials. Controlling the intrinsic stochasticity in RS materials could enable novel electronics applications including multimodal oscillators, true random number generators, probabilistic p-bits, and complex neurons in hardware-level neuromorphic circuits with programmable spiking probability distribution functions.

## Methods

10-nm thick $V_2O_3$ films were grown on $(1\bar{1}02)$ -oriented (R-cut) $Al_2O_3$ substrates via RF magnetron sputtering using a $V_2O_3$ target in a 7.9 mTorr Ar atmosphere and at 640 °C substrate temperature. After the growth, the samples were thermally quenched at a rate of ~90 °C $min^{-1}$, as this helps preserve the correct oxygen stoichiometry and allows for dramatic improvement of electronic properties.[42] (100 nm Au)/(20 nm Ti) electrodes were fabricated in 10×50 $\mu m^2$ two-terminal geometry using standard photolithography techniques and e-beam evaporation. After pristine samples had been characterized, a 1-μm-wide strip bridging the electrodes was irradiated using a 30 keV Ga+ focused ion beam with a fluence of $6.2\cdot10^{14}$ Ga ions/cm$^2$ in a commercial scanning electron microscope.

**Acknowledgements**

This work was supported as part of the Quantum Materials for Energy Efficient Neuromorphic Computing (Q-MEEN-C), an Energy Frontier Research Center funded by the U.S. Department of Energy, Office of Science, Basic Energy Sciences under Award # DE-SC0019273. This work was performed in part at the San Diego Nanotechnology Infrastructure (SDNI) of UCSD, a member of the National Nanotechnology Coordinated Infrastructure, which is supported by the National Science Foundation (Grant ECCS-2025752).

**Author Contributions**

N.G. fabricated the devices, performed the measurements and analyzed the data with input from P.S and I.K.S.. D.A., L.F., and M.R. performed theoretical simulations. I.K.S. supervised the design, execution and data analysis of the experiment. All authors contributed to the interpretation of the results and production of the paper.

**Competing interests**

The authors declare no competing interests.